\documentclass[conference]{IEEEtran}
\IEEEoverridecommandlockouts
\usepackage{cite}
\usepackage{amsmath,amssymb,amsfonts}
\usepackage{algorithmic}
\usepackage{graphicx}
\usepackage{textcomp}
\usepackage{xcolor}
\usepackage{bm}
\usepackage{mathptmx}

\makeatletter
\newcommand{\linebreakand}{%
  \end{@IEEEauthorhalign}
  \hfill\mbox{}\par
  \mbox{}\hfill\begin{@IEEEauthorhalign}
}
\makeatother

\DeclareMathAlphabet{\pazocal}{OMS}{zplm}{m}{n}
\let\Phi\varPhi
\graphicspath{{./figures/}}
\def\BibTeX{{\rm B\kern-.05em{\sc i\kern-.025em b}\kern-.08em
    T\kern-.1667em\lower.7ex\hbox{E}\kern-.125emX}}
\begin{document}

\title{Augmentation of Generalized Multivariable Grid-Forming Control for Power Converters with Cascaded Controllers}

\author{\IEEEauthorblockN{Meng Chen$^1$, Dao Zhou$^1$, Ali Tayyebi$^2$, Eduardo Prieto-Araujo$^3$, Florian D{\"{o}}rfler$^2$, and Frede Blaabjerg$^1$}
\IEEEauthorblockA{$^1$\textit{AAU Energy, Aalborg University}, Aalborg, Denmark}
\IEEEauthorblockA{$^2$\textit{Automatic Control Laboratory, Swiss Federal Institute of Technology (ETH) Zurich}, Zurich, Switzerland}
\IEEEauthorblockA{$^3$\textit{CITCEA-UPC, Technical University of Catalonia}, Barcelona, Spain}
\{mche, zda\}@energy.aau.dk, alit@student.ethz.ch, eduardo.prieto-araujo@upc.edu, dorfler@ethz.ch, fbl@et.aau.dk
\thanks{E Prieto is a lecture of the Serra H{\'{u}}nter programme.}
}

\maketitle

\begin{abstract}
The classic design of grid-forming control strategies for power converters rely on the stringent assumption of the timescale separation between DC and AC states and their corresponding control loops, e.g., AC and DC loops, power and cascaded voltage and current loops, etc. This paper proposes a multi-input multi-output based grid-forming (MIMO-GFM) control for the power converters using a multivariable feedback structure. First, the MIMO-GFM control couples the AC and DC loops by a general multivariable control transfer matrix. Then, the parameters design is transformed into a standard fixed-structure $\pazocal{H}_{\infty}$ synthesis. By this way, all the loops can be tuned simultaneously and optimally without relying on the assumptions of loop decoupling. Therefore, a superior and robust performance can be achieved. Experimental results verify the proposed method.
\end{abstract}


\section{Introduction}
The design of grid-forming controllers is complicated, as there always are many nested loops to be tuned in order to simultaneously achieve multiple objectives, e.g., stability, power regulation, synchronization, etc \cite{Tayyebi2020}. Therefore, several assumptions are usually made in traditional grid-forming control architectures. Firstly, the DC and AC dynamics are supposed to be decoupled \cite{Wu2016}. Secondly, it is assumed that the active and reactive powers can be controlled independently \cite{Liu2020}. Thirdly, the cascaded voltage and current loops are often neglected due to their large bandwidths compared with the power loops, e.g., in low-power converters \cite{Chen2021}. 

The above assumptions simplify the grid-forming control design to several decoupled single-input single-out (SISO) systems, e.g., DC loop, AC power loops, and cascaded loops, which can be separately designed and tuned. When focusing on the AC dynamics, an ideal DC source is typically used in most prevalent grid-forming controllers, e.g., droop control, virtual synchronous generator (VSG), etc \cite{Wu2016,Tayyebi2020}. With respect to the cascaded controllers, the symmetrical optimum criterion and modulus optimum criterion can be used \cite{DArco2013,DArco2015b}. Thereafter, predefined $p$-$f$ and $q$-$V$ relationships are usually used to construct the active and reactive power loops, respectively, where typical designs are based on the classic control methods, e.g., root-locus and frequency analysis \cite{Wu2016,Khajehoddin2019}.

Nevertheless, the aforementioned assumptions can not provide a superior and robust performance. On one hand, the loops decoupling is not always reasonable due to the fact that the low switching frequency of a high-power converter limits the bandwidths of the cascaded controllers. In this context, a complete model has to be used \cite{DArco2013,DArco2015b}. On the other hand, these simplifications limit the bandwidths of the power loops \cite{Chen2019b} and they also discard the useful information in the coupling terms, which have proved to be favorable in \cite{Arghir2018,Shuai2019,tayyebi2020hybrid,Xiong2021}, especially regarding the DC dynamics. Moreover, the manual parameter tuning based on the classic control methods is cumbersome and inconvenient when coping with multiple parameters.

Recently, the $\pazocal{H}_2/\pazocal{H}_{\infty}$ synthesis begins to be used in the grid-forming controller and the VSC-HVDC link design, which, compared with the aforementioned methods, can achieve a multi-objective optimization by simultaneously tuning all the loops \cite{Johnson2018,Fathi2018,Kammer2019,Sanchez2020,Dadjo2022,Huang2020,chen2021generalized}. In \cite{Huang2020}, the power loops and inner voltage loop are considered. However, the important DC dynamics are neglected, which influences the performance of the controllers. A multi-input multi-output based grid-forming (MIMO-GFM) control is proposed in \cite{chen2021generalized} by coupling the DC loop and the AC power loops, which shows a superior and robust performance. Nevertheless, the cascaded controllers are not considered. Therefore, this paper extends the work of \cite{chen2021generalized} by including the cascaded controllers into the fixed-structure $\pazocal{H}_{\infty}$ synthesis. In this way, the following advantages can be achieved.
\begin{enumerate}
	\item The assumptions of bandwiths (or timescales) separation among different loops, e.g., cascaded voltage and current loops, and power loops, can be removed.
	\item The assumptions of loops decoupling, e.g., active and reactive power loops, and DC loop, can be removed.
	\item The parameters can be tuned simultaneously and automatically in order to obtain an optimal performance.
\end{enumerate}

We experimentally validate the superior performance of our design in comparison to a standard VSG control with nested, time-scale separated, and decoupled loops.
 
The remainder of paper is organized as follows. The model of the MIMO-GFM converter with cascaded controllers is built in Section II. In Section III, the formulation of the fixed-structure $\pazocal{H}_{\infty}$ synthesis for the MIMO-GFM converter with cascaded controllers are given. In Section IV, experimental results are shown, and finally, conclusions are given in Section V.

\section{Modeling of MIMO-GFM with Cascaded Controllers}

\begin{figure}[!t]
\centering
\includegraphics[width=\columnwidth]{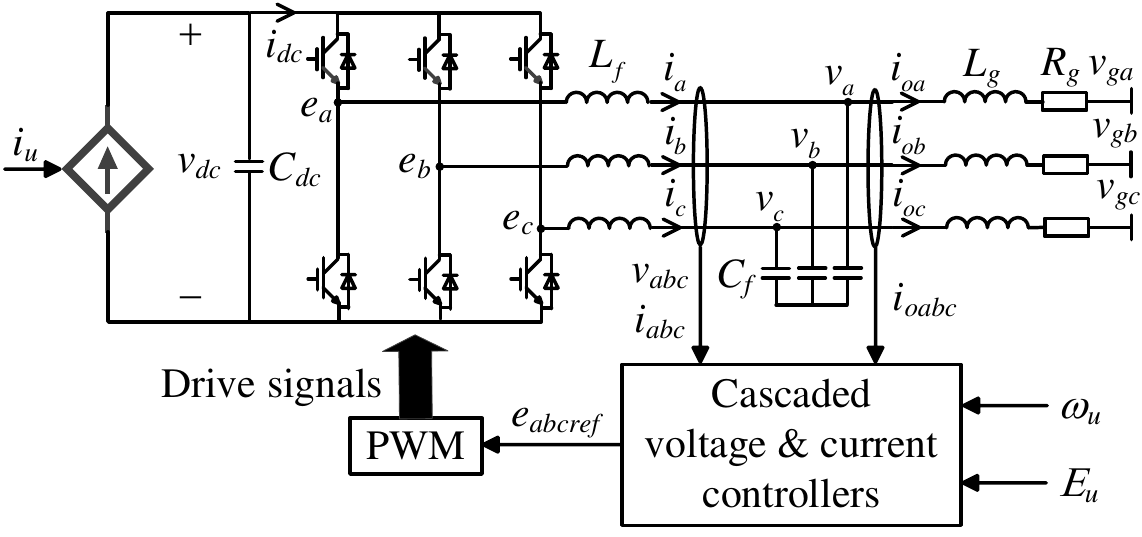}
\caption{DC-AC power converter configuration with low-level cascaded controllers that process the references $i_u$, $\omega_u$, and $E_u$ provided by a high-level grid-forming control strategy.}
\label{VSC}
\end{figure}

Fig. \ref{VSC} shows the topology of the grid-forming converter controlled with cascaded controllers, where the power stage consists of a three-phase converter and an LC filter with $L_f$ and $C_f$ being its inductor and capacitor. In the MIMO-GFM converter, the DC source is represented by a controlled current source $i_u$ in order to consider the dynamics of the DC capacitor $C_{dc}$ and free the assumption of an ideal DC source. Moreover, $\omega_u$ and $E_u$ are the inputs of frequency and voltage to achieve the ability of grid-forming, which are used as the references of the cascaded controllers. Therefore, a MIMO-GFM controller should provide three inputs, i.e., $i_u$, $\omega_u$, and $E_u$ for both the DC and AC loops.

The power stage model of the MIMO-GFM converter of Fig. \ref{VSC}, in the $\omega_u$-defined $d$-$q$ frame can be summarized as 
\begin{align}
\label{id_dynamic}
&\dot i_d=\frac{\omega_b}{L_f}e_d-\frac{\omega_b}{L_f}v_d+\omega_b\omega_ui_q\\
&\dot i_q=\frac{\omega_b}{L_f}e_q-\frac{\omega_b}{L_f}v_q-\omega_b\omega_ui_d\\
&\dot v_d=\frac{\omega_b}{C_f}i_d-\frac{\omega_b}{C_f}i_{od}+\omega_b\omega_uv_q\\
&\dot v_q=\frac{\omega_b}{C_f}i_q-\frac{\omega_b}{C_f}i_{oq}-\omega_b\omega_uv_d\\
&\dot i_{od}=\frac{\omega_b}{L_g}v_d-\frac{\omega_b}{L_g}V_g\cos\delta-\frac{\omega_bR_g}{L_g}i_{od}+\omega_b\omega_ui_{oq}\\
&\dot i_{oq}=\frac{\omega_b}{L_g}v_q+\frac{\omega_b}{L_g}V_g\sin\delta-\frac{\omega_bR_g}{L_g}i_{oq}-\omega_b\omega_ui_{od}\\
\label{vdc_dynamic}
&\dot v_{dc}=\frac{\omega_b}{C_{dc}}i_u-\frac{\omega_b(e_di_d+e_qi_q)}{C_{dc}v_{dc}}
\end{align}
where $e_d$ and $e_q$ are the output voltages, $v_d$ and $v_q$ are the capacitor voltages, $i_d$ and $i_q$ are the inductor currents of the filter, $i_{od}$ and $i_{oq}$ are output currents, $v_{dc}$ is the DC voltage, $V_g$ is grid voltage magnitude, $\omega_b$ is the base value of the angular frequency, $\delta$ is the angle difference between the MIMO-GFM controller frame and the power grid frame, which is defined as
\begin{align}
&\dot\delta=\omega_b\omega_u-\omega_b\omega_g
\end{align}
where $\omega_g$ is the grid frequency. The dynamics of $v_{dc}$ in (\ref{vdc_dynamic}) shows the natural coupling between AC and DC sides.

The cascaded controllers use PI controllers with decoupling and feedforward terms, whose control commands are
\begin{align}
&i_{dref}=k_{pv}(E_u-v_d)+k_{iv}\int_0^t(E_u-v_d)d\tau-C_fv_q+k_{ffi}i_{od}\\
&i_{qref}=k_{pv}(-v_q)+k_{iv}\int_0^t(-v_q)d\tau+C_fv_d+k_{ffi}i_{oq}\\
&e_{dref}=k_{pi}(i_{dref}-i_d)+k_{ii}\int_0^t(i_{dref}-i_d)d\tau-L_fi_q+k_{ffv}v_d\\
&e_{qref}=k_{pi}(i_{qref}-i_q)+k_{ii}\int_0^t(i_{qref}-i_q)d\tau+L_fi_d+k_{ffv}v_q
\end{align}
where $k_{pv}$ and $k_{iv}$ are gains of voltage PI controllers, $k_{pi}$ and $k_{ii}$ are gains of current PI controllers, $k_{ffi}$ and $k_{ffv}$ are gains of feedforward terms of voltage and current controllers, $i_{dref}$ and $i_{qref}$ are current references, $e_{dref}$ and $e_{qref}$ are the references of the converter voltages.

The impact of PWM and sampling delay can be estimated by \cite{DArco2013,DArco2015b}
\begin{align}
&\dot e_d=\frac{1}{1.5T_{sw}}e_{dref}-\frac{1}{1.5T_{sw}}e_d\\
\label{eq_dynamic}
&\dot e_q=\frac{1}{1.5T_{sw}}e_{qref}-\frac{1}{1.5T_{sw}}e_q
\end{align} 
where $T_{sw}$ is the switching cycle. Therefore, the bandwidths of the cascaded controllers is strongly limited by the employed switching frequency. As a result, the cascaded controllers can not be neglected as usual for the power converters with low switching frequencies \cite{DArco2015b}.

For the grid-forming converter, typically five output quantities are considered, i.e., the active and reactive power $p$ and $q$, magnitude of the terminal voltage $V$, the frequency assigned by the grid-forming control $\omega_u$, as well as $v_{dc}$, where $p$, $q$, $V$ are expressed by the state variables as follows:
\begin{align}
&p=v_di_{od}+v_qi_{oq}\\
&q=-v_di_{oq}+v_qi_{od}\\
\label{V_calculation}
&V=\sqrt{v^2_d+v^2_q}
\end{align}

\begin{figure*}[!t]
\centering
\includegraphics[width=\textwidth]{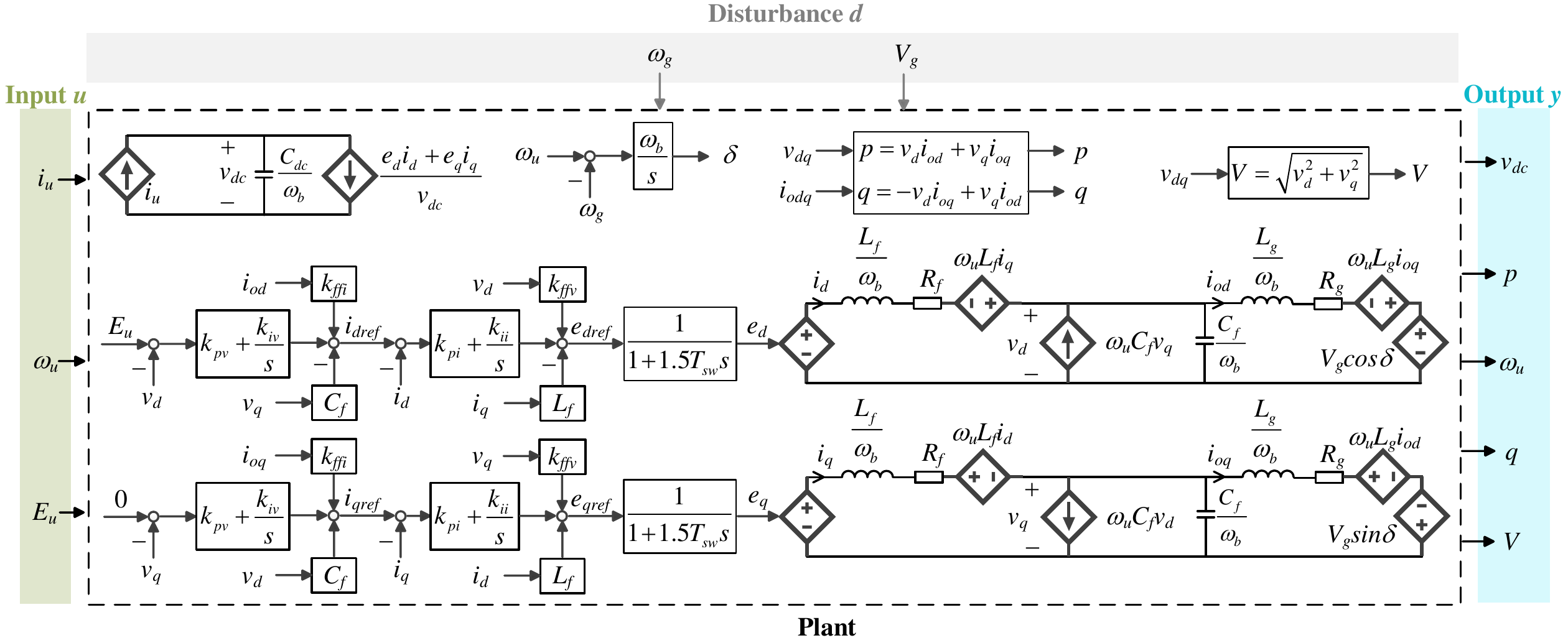}
\caption{MIMO open-loop equivalent circuit of grid-forming converter with cascaded controllers in $d$-$q$ frame.}
\label{equivalent}
\end{figure*}

According to (\ref{id_dynamic})-(\ref{V_calculation}), the MIMO open-loop equivalent model of the grid-forming converter with cascaded controllers is shown in Fig. \ref{equivalent}, which represents a three-input five-output system where the grid voltage is treated as a disturbance.

\begin{figure}[!t]
\centering
\includegraphics[width=0.75\columnwidth]{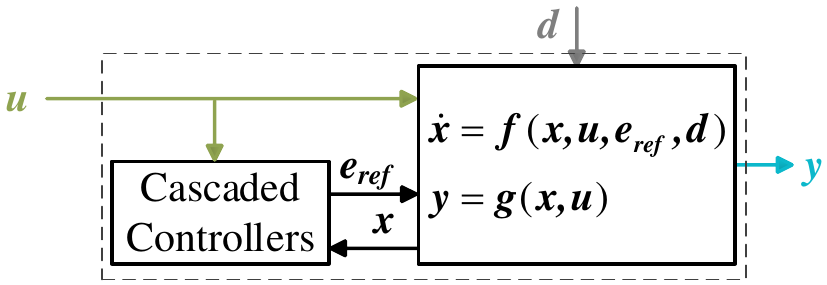}
\caption{Compact MIMO open loop equivalent model of grid-forming converter with cascaded controllers - see Fig. \ref{equivalent} for details.}
\label{open_control}
\end{figure}

\begin{figure}[!t]
	\centering
	\includegraphics[width=\columnwidth]{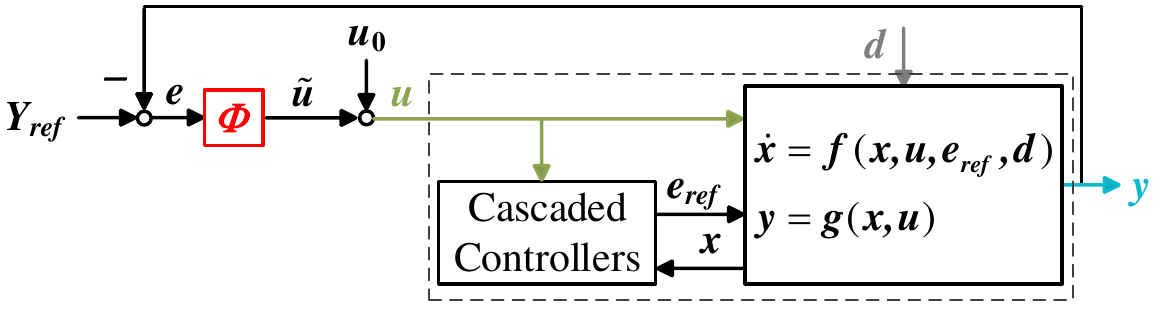}
	\caption{Compact MIMO close loop feedback control of grid-forming converter with cascaded controllers.}
\label{close_control}
\end{figure}

For notational convenience, we define the vectors:
\begin{align}
&\bm x=\left[
	\begin{matrix}
	e_d&e_q&i_d&i_q&v_d&v_q&i_{od}&i_{oq}&v_{dc}&\delta
	\end{matrix}
\right]^T\\
&\bm{e_{ref}}=\left[
	\begin{matrix}
	e_{dref}&e_{qref}
	\end{matrix}
\right]^T,~\bm u=\left[
	\begin{matrix}
	i_u&\omega_u&E_u
	\end{matrix}
\right]^T\\
&\bm y=\left[
	\begin{matrix}
	v_{dc}&p&\omega_u&q&V
	\end{matrix}
\right]^T\\
&\bm d=\left[
	\begin{matrix}
	\omega_g&V_g
	\end{matrix}
\right]^T,
\end{align}
where $\bm x$ is the state vector of the power stage, $\bm u$ is the control vector, $\bm y$ is the output vector, and $\bm d$ is the disturbance vector. Thereafter, the equivalent circuit of Fig. \ref{equivalent} can be abstracted as a compact open-loop model as in Fig. \ref{open_control}. Similar to the control synthesis in \cite{chen2021generalized}, this open-loop model is closed by the multivariable MIMO-GFM feedback control as illustrated in Fig. \ref{close_control}.
\begin{align}
&\bm{u_0}=\left[
	\begin{matrix}
	i_0&\omega_0&E_0
	\end{matrix}
\right]^T\\
&\bm{Y_{ref}}=\left[
	\begin{matrix}
	V_{dcref}&P_{ref}&\omega_g&Q_{ref}&V_{ref}
	\end{matrix}
\right]^T\\
&\bm{\Phi}=\left(\phi_{ij}\right)_{3\times 5},
\end{align}
where $\bm{u_0}$ is the vector of set-points for $\bm u$, $\bm{Y_{ref}}$ is the vector of references for $\bm y$. In addition, $\bm e$ is the vector of error signals. We notice that $\bm{\Phi}=\bm{\Phi}(s)$ is the $3\times5$ control transfer matrix. The performance of the grid-forming controller is therefore grounded on the choices of $\bm{\Phi}$, where a reasonable choice is
\begin{align}
\label{control_matrix}
\bm\Phi=\left[
	\begin{matrix}
	k_{pdc}+k_{idc}/s&k_{12}&0&k_{14}&k_{15}\\
	k_{21}&D_pk_{22}/(s+k_{22})&0&k_{24}&k_{24}/D_q\\
	k_{31}&k_{32}&0&k_{34}/s&\frac{k_{34}/D_q}{s}
	\end{matrix}
\right]
\end{align}
where $D_p$ and $D_q$ are droop coefficients to achieve a proper power sharing in the case of multiple MIMO-GFM converters, the PI controller of $\phi_{11}$ guarantees a zero-error of the DC voltage in the steady-state, the low-pass filter of $\phi_{22}$ provides inertia characteristics like in a synchronous generator, and the third column with zeros avoids the dependence on $\omega_g$ and, therefore, realize a complete phase-locked-loop-free (PLL-free) system. It is noted that $\bm{\Phi}$ in (\ref{control_matrix}) couples all the DC and AC power loops in a single monolithic controller. 

\section{$\pazocal{H}_{\infty}$ Synthesis of MIMO-GFM with Cascaded Controllers}

As shown in (\ref{control_matrix}), unlike perceiving the grid-forming converter as several decoupled SISO systems \cite{Wu2016,Liu2020,Chen2021}, the MIMO-GFM controller accounts for the natural system couplings and thus highly improves the performance of the power converter by assigning simple proportional controllers rather than high-order terms (e.g., damping terms based on PLL or high-order filters). Thereafter, another problem is how to choose the parameters for an optimal performance. Classic methods are manual tuning based on root-locus or frequency analysis, which may be very inconvenient when simultaneously selecting multiple parameters. Therefore, the bandwidths of the cascaded voltage and current loops as well as power loops are usually chosen in different timescales. Then they are designed separately, which, nevertheless, has several shortages. First, the loops' decoupling between DC and AC sides is not true for the MIMO-GFM controller. Second, the bandwidths of the cascaded voltage and current loops may not be high enough for a high-power converter with low switching frequency (e.g., 2 kHz). Third, the performance may be limited in order to keep the bandwidths separation. To cope with these issues, in what follows we transform the parameters design of the MIMO-GFM controller with cascaded controllers into an $\pazocal{H}_{\infty}$ synthesis. As a result, all the nested loops are simultaneously tuned to obtain an optimal performance.

\begin{figure}[!t]
	\centering
	\includegraphics[width=\columnwidth]{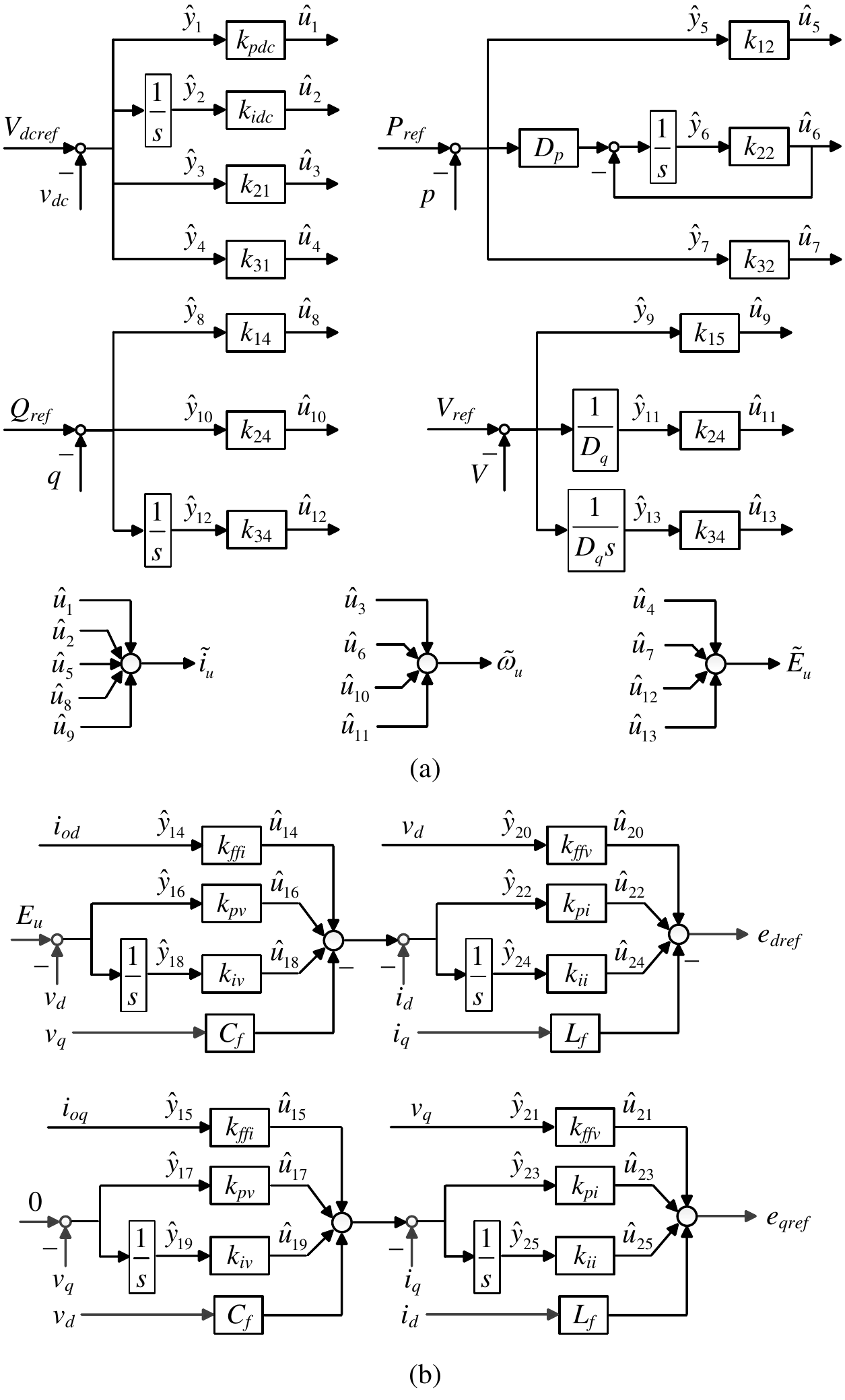}
	\caption{Block diagram of used (a) control transfer matrix and (b) cascaded controllers.}
	\label{h_infinity_formulation}
\end{figure}

\begin{figure}[!t]
	\centering
	\includegraphics[width=0.4\columnwidth]{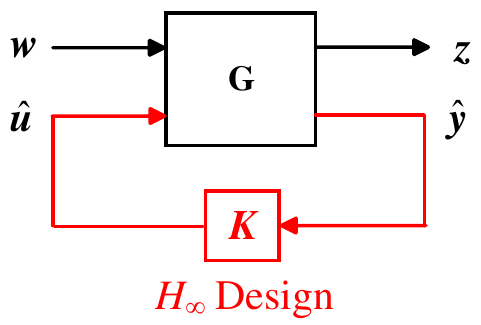}
	\caption{Block diagram of grid-forming converter with cascaded controllers in linear fractional transformation.}
	\label{LFT}
\end{figure}

To this end, the tuned parameters in both the control transfer matrix $\bm{\Phi}$ and cascaded controllers should be separated as shown in Fig. \ref{h_infinity_formulation}. Two intermediate vectors $\bm{\hat y}$ and $\bm{\hat u}$ are introduced and have the following relationship
\begin{align}
\hat{\bm u}=&diag(k_{pdc},k_{idc},k_{21},k_{31},k_{12},k_{22},k_{32},k_{14},k_{15},k_{24}\bm{I_2},k_{34}\bm{I_2},\notag\\
&k_{ffi}\bm{I_2},k_{pv}\bm{I_2},k_{iv}\bm{I_2},k_{ffv}\bm{I_2},k_{pi}\bm{I_2},k_{ii}\bm{I_2})\hat{\bm y}\notag\\
=&\bm K\hat{\bm y}
\end{align}
where the static gain $\bm K$ contains all the parameters to be tuned. Thereafter, the standard structure of linear fractional transformation for $\pazocal{H}_{\infty}$ synthesis can be derived as shown in Fig. \ref{LFT} by collapsing the MIMO-GFM converter of Fig. \ref{close_control} (except for $\bm K$) into \textbf{G}. As shown in Fig. \ref{LFT}, $\bm w$ and $\bm z$ are the defined disturbance inputs and performance outputs for the $\pazocal{H}_{\infty}$ synthesis, which can be chosen according to the control objectives.

As an example, this paper considers a multi-objective problem with both active power regulation and synchronization, where $\bm w$ and $\bm z$ are chosen as
\begin{align}
\label{w}
&\bm w=\left[
	\begin{matrix}
	\Delta{P_{ref}}&\Delta\omega_g
	\end{matrix}
\right]^T\\
&\bm z= \left[
	\begin{matrix}
	\Delta{P_{ref}}-\frac{1}{D_p}\Delta\omega_g-\Delta{p}&\Delta{p}
	\end{matrix}
\right]^T.
\end{align}

We use $T_{ij}$ to represent the transfer function from $w_j$ to $z_i$. In the $\pazocal{H}_{\infty}$ synthesis, some weighting functions should be designed to limit the focused $T_{ij}$ in order to obtain a favorable performance of the MIMO-GFM converter.

Specifically, $T_{11}$ reflects the error of power tracking according to the droop characteristics. In order to have quick dynamics and small error, its weighting function is chosen as
\begin{align}
W_{11} = \frac{s+8}{s+0.0008}
\end{align}

Similarly, $T_{21}$ reflects the dynamics of the output active power. In order to limit the high-frequency disturbance, its weighting function is chosen as
\begin{align}
W_{21} = \frac{1/80s+1}{1/8000s+1}
\end{align}

Last, $T_{12}$ reflects the power regulation responding to the disturbance of the grid frequency. In order to have quick dynamics and small error, its weighting function is chosen as
\begin{align}
\label{w12}
W_{12} = \frac{s+6}{100s+0.0006}
\end{align}

It is worth mentioning that the above equations of (\ref{w})-(\ref{w12}) only show an example of how to use the $\pazocal{H}_{\infty}$ synthesis to tune the MIMO-GFM converter with the cascaded controllers, where they can be changed according to the required performance. Meanwhile, other disturbance inputs, performance outputs, and weighting functions (e.g., for reactive power and voltage) can also be designed analogously if necessary. Thereafter, the static gain $\bm K$ can be obtained (e.g., by instructor $hifstruct$ in Matlab) from
\begin{align}
\label{h_infinity}
\min_{\bm K}||diag(W_{ij}(s)T_{ij}(s))||_\infty.
\end{align}

\section{Experimental Validation}
\begin{figure}[!t]
\centering
\includegraphics[width=\columnwidth]{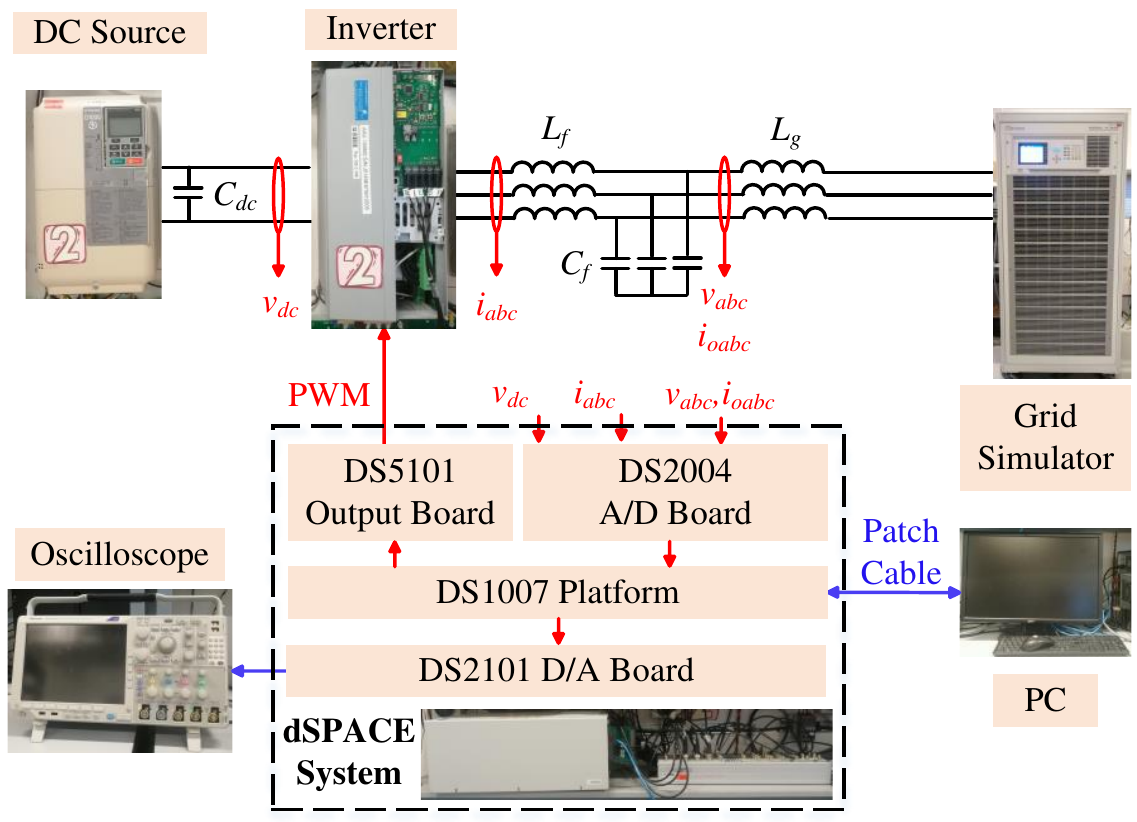}
\caption{Experimental configuration of MIMO-GFM converter with cascaded controllers.}
\label{setup}
\vspace{-10pt}
\end{figure}

In order to show the benefits of the $\pazocal{H}_{\infty}$-tuned MIMO-GFM converter with cascaded controllers, the experimental results will be presented, where the configuration of the setup is shown in Fig. \ref{setup}, and the key parameters are given in Table \ref{parameter}.

\begin{table}[!t]
	\renewcommand{\arraystretch}{1.3}
	\caption{Parameters of Experimental Setups}
	\centering
	\label{parameter}
	\resizebox{\columnwidth}{!}{
		\begin{tabular}{c l c}
			\hline\hline \\[-3mm]
			Symbol & Description & Value  \\ \hline
			$\omega_n$  & Nominal frequency & $100\pi$ rad/s \\
			$S_n $ & Nominal power &  5 kW  \\ 
			$V_n $ & Nominal line-to-line RMS voltage & 380 V \\
			$f_{sw}$ & Switching frequency & 10 kHz \\
			$\omega_g $ & Grid frequency & $100\pi$ rad/s (1 p.u.)  \\
			$V_g $ & Line-to-line RMS grid voltage & 380 V (1 p.u.) \\
			$L_g$  & Line inductor & 8 mH (0.087 p.u.)\\
			$C_f$ & Filter capacitor & 5 $\mu$F (0.0454 p.u.)\\   
			$L_f$ & Filter inductor & 3 mH (0.0326 p.u.)\\
			$C_{dc}$ & DC capacitor & 500 $\mu$F \\ 
			$D_p$ & Droop coefficient of $P$-$f$ regulation & 0.01 p.u. \\ 
			$D_q$ &Droop coefficient of $Q$-$V$ regulation & 0.05 p.u.\\
			$P_{ref}$ & Active power reference & 0.5 p.u.\\
			$Q_{ref}$ & Reactive power reference & 0 p.u.\\
			$V_{ref}$ & Voltage magnitude reference & 1 p.u.\\
			$V_{dcref}$ &DC voltage reference & 700 V\\[1.4ex]
			\hline\hline
		\end{tabular}
	}
\end{table}

\subsection{Classic methods of parameters design}

\begin{figure}[!t]
	\centering
	\includegraphics[width=0.9\columnwidth]{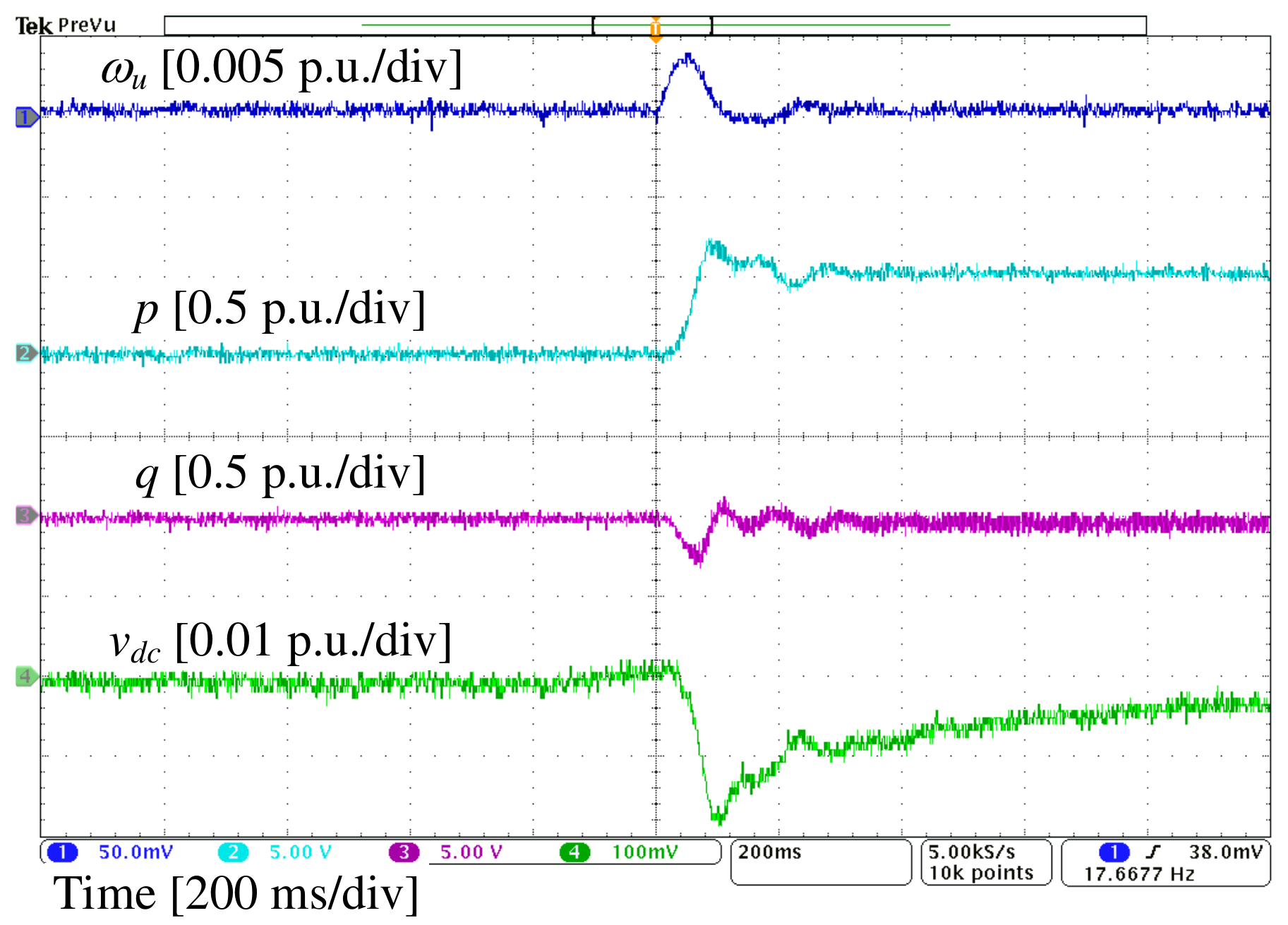}
	\caption{Experimental results of traditional VSG when $P_{ref}$ steps from 0.5 p.u. to 1 p.u.}
	\label{experiment_pref_tradition}
\end{figure}

\begin{figure}[!t]
	\centering
	\includegraphics[width=0.9\columnwidth]{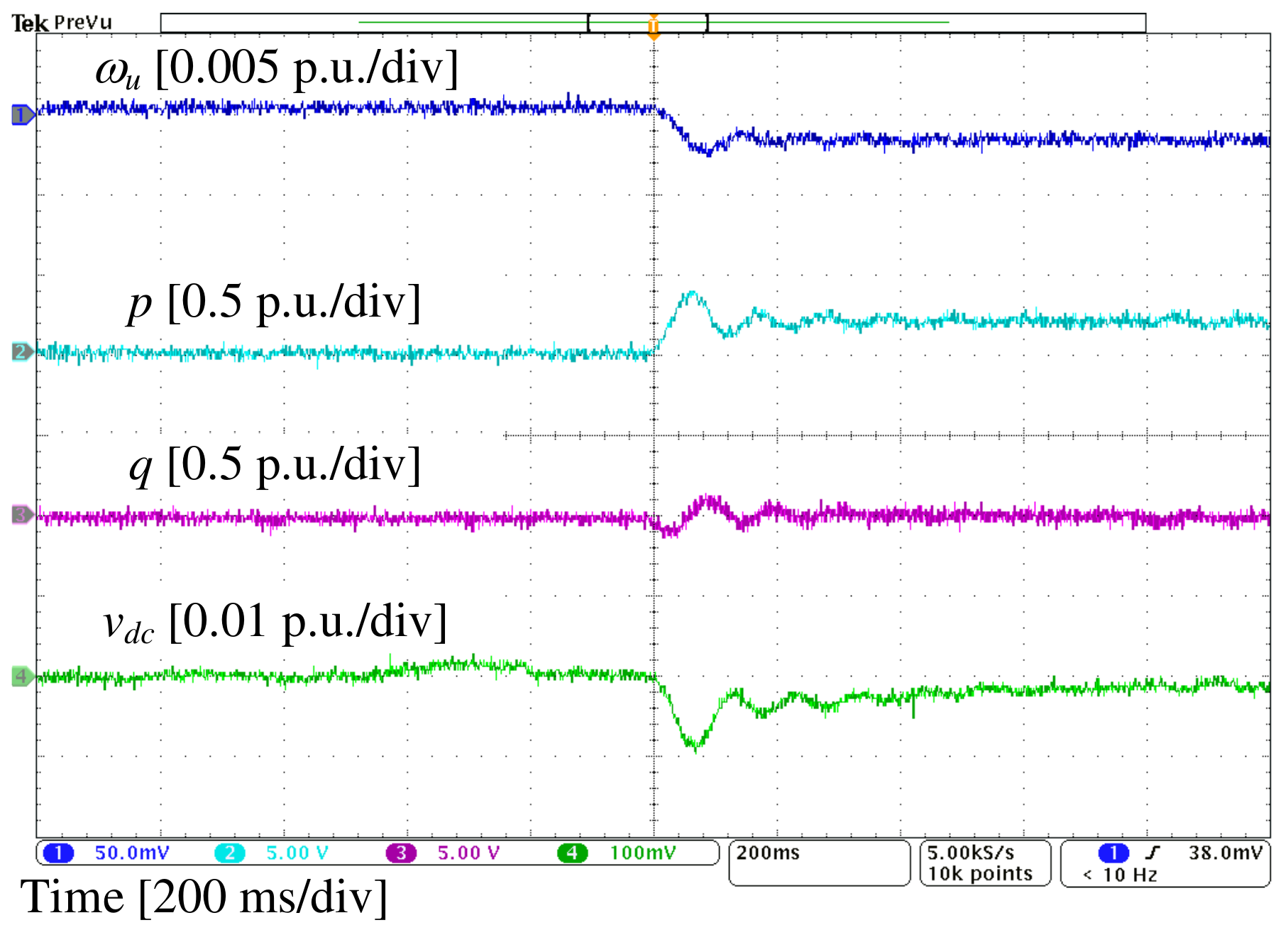}
	\caption{Experimental results of traditional VSG when $\omega_g$ decreases from 50 Hz to 49.9 Hz}
	\label{experiment_wg_tradition}
\end{figure}

Before giving the results of the proposed $\pazocal{H}_{\infty}$-tuned MIMO-GFM converter with cascaded controllers, the results with the classic methods for the popular virtual synchronous generator (VSG) are given as a comparison of performance in this section. The cascaded controllers are designed by the symmetrical optimum criterion and modulus optimum criterion as shown in \cite{DArco2015b}. Furthermore, the power loops are designed based on the conclusions in \cite{Chen2019b}. As the results, the derived parameters are listed in Table \ref{designed_parameter}. Fig. \ref{experiment_pref_tradition} and Fig. \ref{experiment_wg_tradition} show the waveform responding to $P_{ref}$ stepping from 0.5 p.u. to 1 p.u. and $\omega_g$ decreasing from 50 Hz to 49.9 Hz, respectively. The dynamics present a stable system but with some undesired oscillations and overshoots.

\begin{table}[!t]
	\renewcommand{\arraystretch}{1.3}
	\caption{Parameters Designed by Different Methods}
	\centering
	\label{designed_parameter}
	\resizebox{\columnwidth}{!}{
		\begin{tabular}{c c c}
			\hline\hline \\[-3mm]
			Parameters & Traditional VSG \cite{DArco2015b,Chen2019b} & \begin{tabular}{@{}c@{}}$\pazocal{H}_{\infty}$-tuned\\MIMO-GFM converter  \end{tabular}  \\ \hline
			$k_{pi}$  & 0.3463 & 0.1371 \\
			$k_{ii}$ & 4.6168 &  16.7853  \\ 
			$k_{ffv}$ & 1 & 0.1223 \\
			$k_{pv}$ & 0.5982 & 0.7738 \\
			$k_{iv}$ & 1026.5 & 1136 \\
			$k_{ffi}$ & 0 & -0.1481 \\
			$k_{21}$  & 0 & -0.1956\\
			$k_{22}$& 30 & 45.1987\\   
			$k_{24}$& 0 & -0.0458\\
			$k_{31}$ & 0 & -0.5115 \\ 
			$k_{32}$& 0 & 0.0167 \\ 
			$k_{34}$&0.1 & 0.624\\[1.4ex]
			\hline\hline
		\end{tabular}
	}
\end{table}

\subsection{$\pazocal{H}_{\infty}$ Synthesis of parameters design}

\begin{figure}[!t]
	\centering
	\includegraphics[width=0.9\columnwidth]{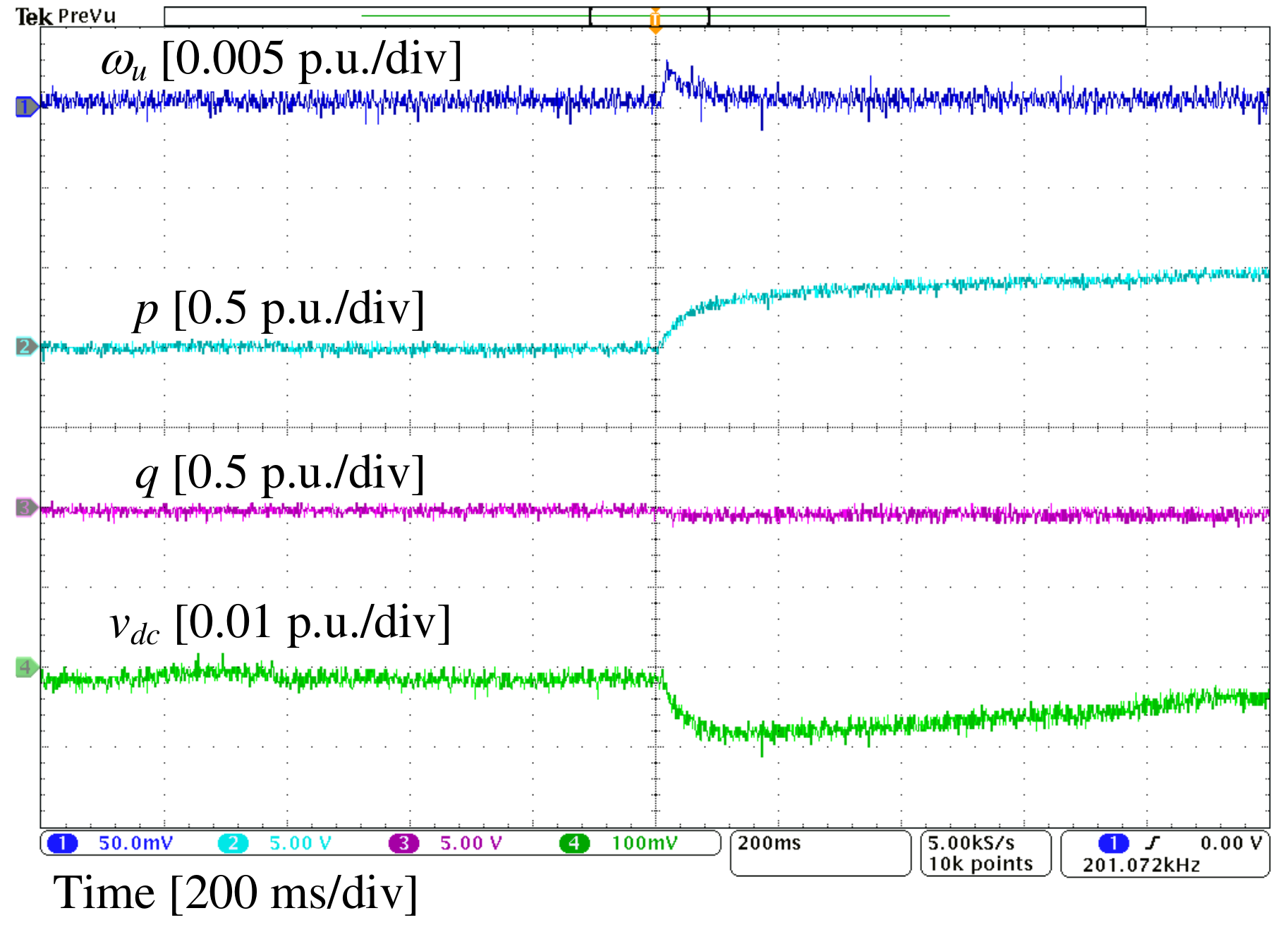}
	\caption{Experimental results of $\pazocal{H}_{\infty}$-tuned MIMO-GFM converter when $P_{ref}$ steps from 0.5 p.u. to 1 p.u.}
	\label{experiment_pref_hinfinity}
\end{figure}

\begin{figure}[!t]
	\centering
	\includegraphics[width=0.9\columnwidth]{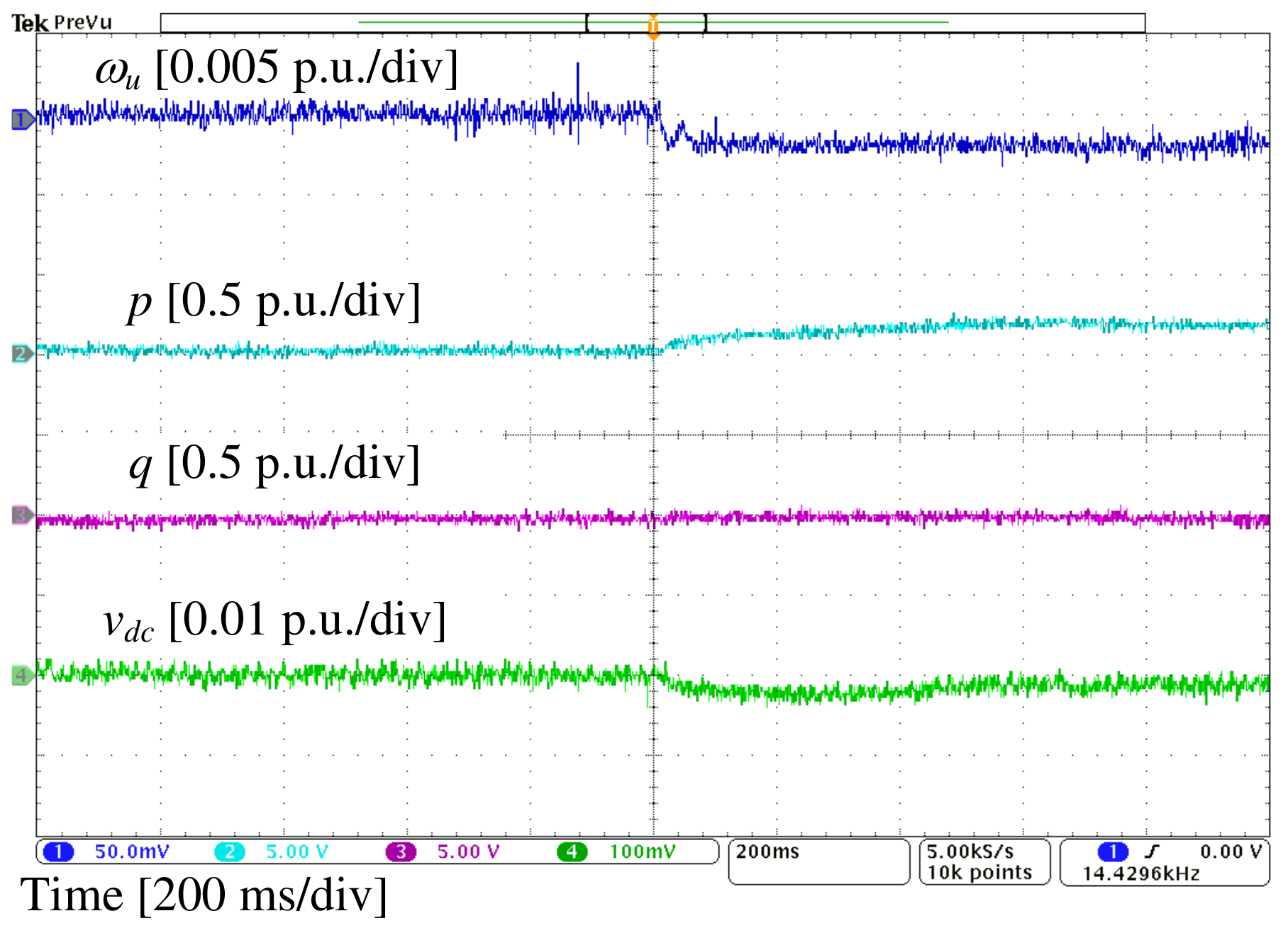}
	\caption{Experimental results of $\pazocal{H}_{\infty}$-tuned MIMO-GFM converter when $\omega_g$ decreases from 50 Hz to 49.9 Hz}
	\label{experiment_wg_hinfinity}
\end{figure}

By solving (\ref{h_infinity}) with the parameters of the VSG as initial values, the parameters of the MIMO-GFM converter with cascaded controllers can be derived by $\pazocal{H}_{\infty}$ Synthesis, where the results are also listed in Table \ref{designed_parameter}. It should be mentioned that $k_{12}$, $k_{14}$, and $k_{15}$ are kept at 0 due to the fact that the DC control of the experimental setup is unchangeable. Fig. \ref{experiment_pref_hinfinity} and Fig. \ref{experiment_wg_hinfinity} show the waveforms of the same test cases as the traditional VSG in Section IV-A. It is observed that the MIMO-GFM converter tuned by the $\pazocal{H}_{\infty}$ synthesis has much smoother dynamics with almost no overshoot. Importantly, the design of the $\pazocal{H}_{\infty}$-tuned MIMO-GFM converter is only determined by the chosen disturbance inputs $\bm w$, evaluation outputs $\bm z$, and weighting functions $\bm W$, which are dependent on the required performance but not on assumptions of loops decoupling and bandwidths separation.

\section{Conclusion}
This paper proposes a design method of MIMO-GFM converter with cascaded controllers. A control transfer matrix based on the multivariable feedback is used to couple the DC and AC power loops. Thereafter, it shows that the parameters design can be transformed into a fix-structured $\pazocal{H}_{\infty}$ optimization problem, where all the parameters including the ones of the cascaded controllers can be tuned simultaneously. By this way, the assumptions of loops decoupling and bandwidths separation are freed and superior performances are obtained, which are illustrated by some design and experimental tests.
\bibliographystyle{IEEEtran}
\bibliography{IEEEabrv,ECCEAsia}
\end{document}